\documentclass[letterpaper]{natureprintstyle}

\usepackage{amsfonts}

\usepackage{eurosym}

\usepackage{graphicx,epsfig}
\usepackage{amsmath,amssymb,bbm}
\usepackage{color}
\usepackage[usenames,dvipsnames]{xcolor}
\usepackage[colorlinks=true,linkcolor=Red,citecolor=Green,urlcolor=Black,linktoc=page]{hyperref}
\usepackage{multirow}
\usepackage[varg]{txfonts}

\usepackage[normalem]{ulem} 

\usepackage{fancyhdr}
\usepackage{mathtools}
\usepackage[caption=false]{subfig}
\usepackage{balance}
\usepackage{float}
\usepackage{flushend}

\DeclareCaptionLabelSeparator{csdel}{\,$\vert$\,}
\usepackage[figurename=Fig.,labelfont=bf,labelsep=csdel]{caption}

\usepackage[normalem]{ulem} 
\usepackage[backend=biber,style=alphabetic,style=nature]{biblatex}  
\renewcommand{\cite}{\autocite}  
\AtBeginBibliography{\small} 
\AtBeginBibliography{\vspace*{-5pt}} 

\def\subinrm#1{\sb{\rm#1}}                                                       
{\catcode`\_=13 \global\let_=\subinrm}                                           
\mathcode`_="8000                                                                
\def\upsubscripts{\catcode`\_=12 }                                               

\upsubscripts

\begin{document}

\title{Topological polarization networking in uniaxial ferroelectrics}

\author{Y.\,Tikhonov$^{1,2}$, J.\,R.\,Maguire$^{3}$, C.\,J.\,McCluskey$^{3}$, J.\,P.\,V.\,McConville$^{3}$, A.\,Kumar$^{3}$, D.\,Meier$^{4}$, A.\,Razumnaya$^{2}$, J.\,M.\,Gregg$^{3}$, A.\,Gruverman$^{5}$, V.\,M.\,Vinokur$^{6,7,1*}$ \& I.\,Luk'yanchuk$^{1}$}

\date{Discovery of topological polarization textures has put ferroelectrics at the frontier of topological matter science. 
High-symmetry ferroelectric oxide materials allowing for freedom of the polarization vector rotation offer a fertile ground for emergent topological polar formations, like vortices, skyrmions, merons, and Hopfions.  
It has been commonly accepted that uniaxial ferroelectrics do not belong in the topological universe because strong anisotropy imposes insurmountable energy barriers for topological excitations. 
Here we show that uniaxial ferroelectrics provide unique opportunity for the formation of topological polarization networks comprising branching intertwined domains with opposite counterflowing polarization. We report that 
they host the topological state of matter: a crisscrossing structure of topologically protected colliding head-to-head and tail-to-tail polarization domains, which for decades has been considered impossible from the electrostatic viewpoint. 
The domain wall interfacing the counterflowing domains is a multiconnected surface, propagating through the whole volume of the ferroelectric.}
\maketitle
\upsubscripts
\thispagestyle{fancy} 
\lfoot{\parbox{\textwidth}{ \vspace{0.0cm}
 \rule{\textwidth}{0.2pt}
\hspace{-0.2cm} \textsf{\scalefont{0.80}
    $^1$University of Picardie, Laboratory of Condensed Matter Physics, Amiens, 80039, France;
    $^2$Faculty of Physics, Southern Federal University, 5 Zorge str., 344090 Rostov-on-Don, Russia; 
    $^3$Centre for Nanostructured Media, School of Mathematics and Physics, Queen's University Belfast, Belfast, BT7 1NN UK;
     $^4$Department of Physics and Astronomy, University of Nebraska–Lincoln, Lincoln, NE, 68588, USA; 
     $^5$Department of Materials Science and Engineering, Norwegian University of Science and Technology (NTNU) Trondheim 7491, Norway;
    $^6$Terra Quantum AG, St. Gallerstrasse 16A, CH-9400 Rorschach, Switzerland;
   $^7$Physics Department, City College of the City University of New York, 160 Convent Ave, New York, NY 10031, USA;
    $^*$The correspondence should be sent to vv@terraquantum.swiss
} 
\vspace{-0.2cm}
\begin{center}{\scalefont{0.87} \thepage}\end{center}}} \cfoot{}

\vspace{1cm}

 The basic property of a ferroelectric is that however complex, the configuration of the polarization distribution has to maintain its internal electric neutrality. The neutrality implies that the polarization vector field is divergenceless, hence similar to the velocity field in incompressible liquids. This similarity manifests in, for example, the recently discovered nanoscale vortex-like topological excitations in ferroelectrics\,\cite{Yadav2016} and whirls in water streams. 
 The polarization vector is free to rotate 
 in the common pseudocubic perovskite ferroelectrics, which leads to a plethora of the observed topological excitations\,\cite{Lahoche2008,Yadav2016,Nahas2015,Das2019,Wang2020,Lukyanchuk2020}. 
 
 In uniaxial ferroelectrics, in contrast, the only established topological excitations are domain walls (DWs), because the spontaneous polarization has only two stable orientations aligned with the anisotropy polar axis.
 Moreover, the electroneutrality condition imposes strong additional restrictions.
 Namely, the inclination of the DWs with respect to 
 the polar axis, 
 leads to head-to-head (H-H) or tail-to-tail (T-T) 
 collisions of polarization vectors belonging to the 
 oppositely-oriented domains, adjacent to the DWs from different sides, hence to the emergence of the 
 bound charges\,\cite{Tagantsev2010book}. To avoid the charging, domain walls need to align 
 with the polarization direction.
 As a result, domains usually spread across the entire sample from the lower to upper surface acquiring a shape of cylinders or stripes. 
 At the same time, the long but finite cigar-like domains have been 
 observed in several 
ferroelectrics\,\cite{Chynoweth1960}. 
Deviations from the electroneutrality at the loci of domains terminations were attributed to the screening by free charge carriers, compensating the bound charges\,\cite{Landauer1957}. 

Here we demonstrate that uniaxial ferroelectrics can, in fact, harbor chargeless polarization  configurations of arbitrary complexity. 
We establish that following the same fundamental laws of topological hydrodynamics  as topological excitations 
in cubic perovskite ferroelectrics, the domains arrange themselves into an
ample-structured topological network of counterflowing mutually avoiding polarization fluxes. 


The electroneutrality of ferroelectrics follows from the requirement of minimizing the electrostatic energy associated with the depolarisation fields. 
To achieve it, 
the polarization distribution tends to assume a structure with vanishing bound charges so that $\mathrm{div}\,\textbf{P} = 0$. The divergenceless nature of the polarization field is a 
basic condition defining the physics of the spatially nonuniform ferroelectricity and resulting in the topological excitations. Therefore, by their topological characteristics, the polarization lines are identical to the streamlines of an ideal incompressible liquid, which enables us to employ the methods of the topological hydrodynamics\cite{Arnold1999}. 

The insight is gained by comparing encountering polarization fluxes with the meeting of countercurrents in incompressible liquid that slightly deviate from their original paths to avoid the clash. 
The counterflowing streams in the liquid correspond thus to the oppositely oriented polarization domains which bend to avoid the H-H and T-T collisions.
Then the neutrality is provided by the topological entwining of the encountered polarization line streams, separated by the single path-connected DW extended through the system. The energy cost related to the domain bending is far below the expense of the electrostatic energy associated with the onset of uncompensated bound charges. 

\begin{figure*}[h!]
\center
\includegraphics [width=19cm] {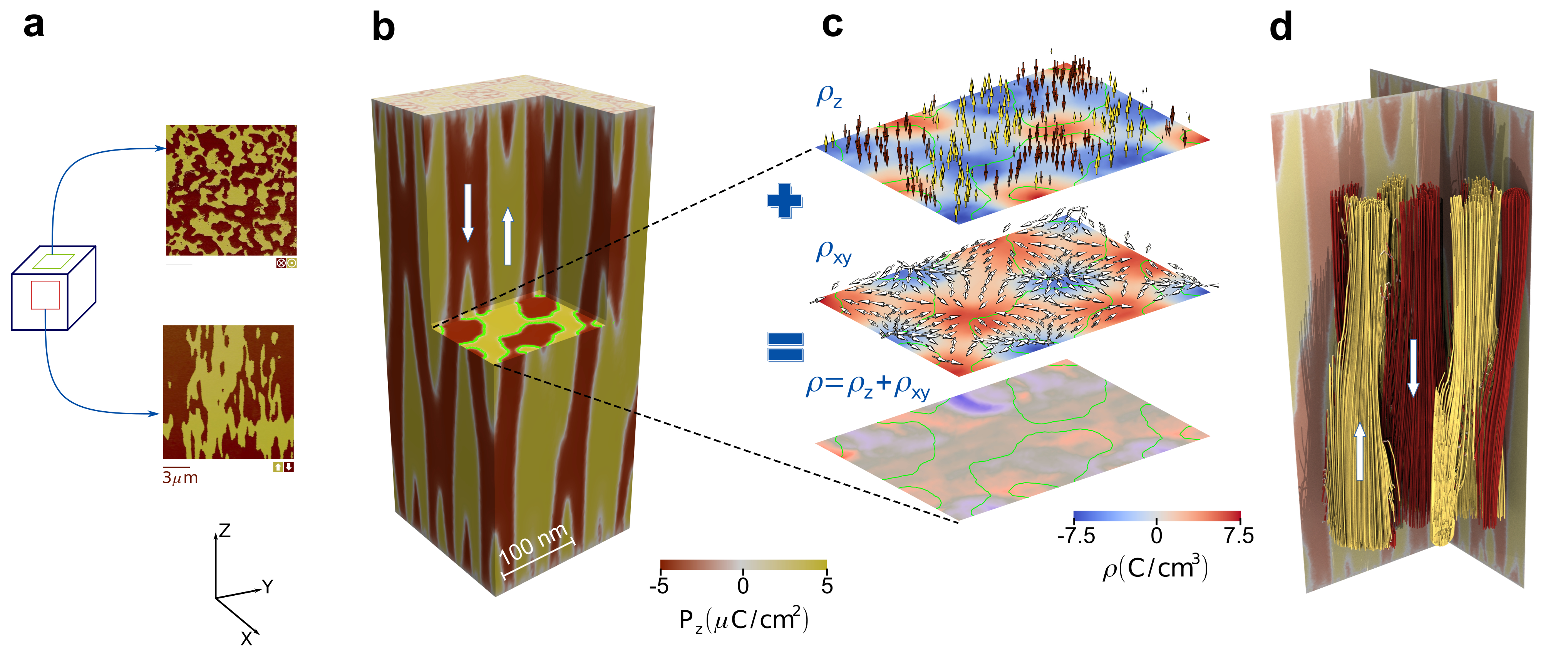}
\caption{ \textbf{The domain networking structure.} 
\textbf{a,}\,Lateral and vertical PFM data from the polar and non polar face of lead germanate, respectively. 
\textbf{b,}\,An inner-corner cut of the rectangular uniaxial PGO sample.
The panel displays the simulated structure of interpenetrating polarization domains with the respective up- and down- polarization directions shown by white arrows. The magnitude of the z-component of the polarization, $P_z$, is quantified by the colour legend. 
 The upper and bottom near-surface regions host Landau fractal multidomain structure arising to compensate surface bound charges. The green contours depict the cross sections of the domain walls.
%
\textbf{c,}\,The charge map of the horizontal cross-section area shown in panel \textbf{b}. The upper plane displays the spontaneous polarization charge $\rho_z=-\partial P_z$, the up– and down- arrows in the upper plane showing the local directions of the $P_z$ polarization. Middle plane displays the charge $\rho_{xy}=-\partial_x P_x - \partial_y P_y$ induced by the transverse polarization whose local directions are shown by the in-plane arrows. The bottom plane displays the total resulting charge $\rho=-\mathrm{div}\,\mathbf{P}$ and demonstrates that $\rho_{xy}$ effectively screens $\rho_z$ so that the total charge, $\rho$, is vanishingly small in the vicinity of the inclined DWs.
 \textbf{d,}\,Visualization of the complex network of polarization lines formed by the counter-flowing polarization fluxes corresponding to ``up" and ``down" domains.
}
\label{FigDomain}
\end{figure*}

To reveal the topology of the encountering polarization fluxes we explore the domain structure in the ferroelectric, lead germanate, 
Pb$_5$Ge$_3$O$_{11}$ (PGO), which is a uniaxial ferroelectric featuring a three-fold polar axis along the [001] crystallographic axis\cite{Iwasaki1971,Nanamatsu1971,Iwasaki1972}. This material exhibits a remnant polarization of P = 4.8 $\mu$C/cm$^2$, which develops through a second order phase transition at $T_c=177$\,$^\circ$C where the space group symmetry changes from P (a paraelectric phase) to P$_3$ (a ferroelectric phase). A single crystal of PGO is grown by the Czochralski method with the seed crystal being pulled from the melt along the [001] direction. The samples are oriented and cut with the diamond saw to produce the parallel-plane plates of several square millimeters in area and with the thickness of about 0.5\,mm. The sample plates are polished with the diamond paste to create surfaces with an optical quality finish. Cutting along the (001) and (100) planes yields the samples with the polar (z-cut) and non-polar (x-cut) surfaces.

The as-grown domain structure in the PGO is mapped using the piezoresponse force microscopy (PFM)\,\cite{Gruverman2019} on the x- and z-cut surfaces as shown in Fig.\,1a. The measurements were taken 
using a Veeco Dimension 3100 AFM system with a Nanoscope IIIa controller and external lock-in amplifier. An AC bias of amplitude 5V and frequency 20kHz was applied to a Pt-Ir coated silicon AFM tip (Nanosensors, PPP-EFM) while in contact with the sample surface. 
The out-of-plane polarization distribution was mapped by monitoring vertical displacement of the cantilever when measurements were performed on the polar face of the crystal, and the in-plane polarization map was obtained by monitoring the lateral PFM signal while scanning the non-polar surface. 
The PFM image acquired on the non-polar surface shows a pattern of domains elongated along the [001] polar axis. The three-dimensional illustration of the as-grown domain arrangement displays the difference in characteristic domain size along the different
crystallographic directions, reflecting the uniaxial nature of ferroelectricity of the material. The PFM image obtained on the polar surface displays an as-grown pattern of irregular domains with antiparallel out-of-plane polarization directions.
In addition to the neutral 180$^\circ$ domain walls, there are also sections of the domain boundaries that correspond to the H-H and T-T arrangements. 
This observation corroborates the previously reported domain walls features\,\cite{Bak2020,Shur1989} by showing that PGO is an unusual example of a uniaxial proper ferroelectric that readily develops uncharged T-T and H-H domain walls.
\begin{figure*}[h!]
\center
\includegraphics [width=18cm] {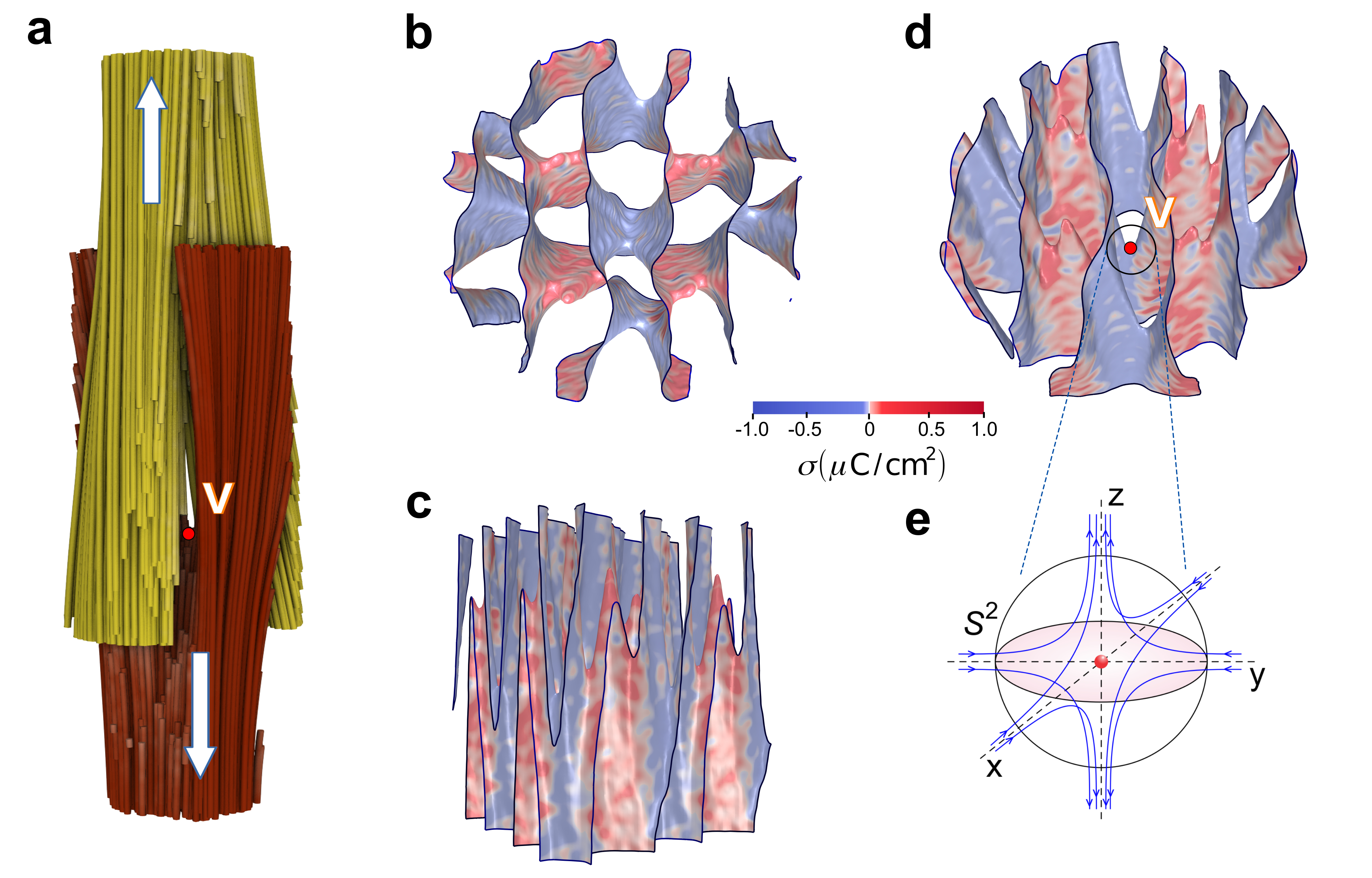}
\caption{ \textbf{The domain wall's topological structure.} 
\textbf{a,} The collision and subsequent splitting of the polarization streams in point $\mathbb{V}$  result in the scissoring pattern.  
\textbf{b,}\,The top view of the domain wall. 
\textbf{c,} The side view of the domain wall.
\textbf{d,}\,The isometric projection of the domain wall. The point V  corresponds a scissoring of the streams in \textbf{a}.
Panels \textbf{b--d} demonstrate that domain wall is a single path-connected surface separating the channels carrying the continuous oncoming streams of polarization fluxes piercing the system. 
The color map visualizes the effective surface charge density, $\sigma$, quantified by the the colour legend. Importantly, the  value of $\sigma$ is substantially less than the polarization reversal $\Delta P_z \simeq \pm 10 \mu C/cm^2$ at the head-to-head and tail-to-tail domain walls because of the lateral screening effect.
\textbf{e,}\, The enlarged vicinity of the saddle singular point $\mathbb{V}$. The pink disc stands for the piece of the plain tangent to the saddle surface in the point $\mathbb{V}$. The blue lines depict the polarization flow in the vicinity of the singularity. 
}
\label{FigDomTopology}
\end{figure*}

To gain further insight into the internal domain structure in the PGO, we perform a quantitative study of the polarization distribution, using phase-field modeling of the bumping polarization fluxes, 
see Methods for details. 
The simulations are based on the minimization of the Ginzburg-Landau functional, F=$\int f dV$  with respect to the magnitude of the z-directed spontaneous polarization, $P$, playing the role of the order parameter describing the uniaxial ferroelectric material and to the electric potential, $\varphi$, determining the electric field $\mathbf{E}=-\nabla \varphi$.
The free energy density is
\begin{gather}
f=\frac{1}{2}A(T-T_c)P^{2}+\frac{b}{4}P^{4}+\frac{c}{6}P^{6}+\frac{G}{2}\left( \nabla P\right) ^{2}+P\partial _{z}\varphi -\frac{1}{2}\varepsilon _{0}\varepsilon _{i}\left( \nabla \varphi \right) ^{2}\,.
\label{GL}
\end{gather}
The total polarization including also a non-critical part induced by the depolarization field is given by $\mathbf{P}=P\mathbf{z}+\varepsilon_0 \varepsilon_i \mathbf{E}$, where we assume that the background dielectric constant $\varepsilon_i$ is isotropic. The numerical values of the coefficients in Eq.\,(\ref{GL}) are obtained from fitting the experimental data reported in the preceding studies of the PGO\,\cite{Iwasaki1972}, 
see Supplementary Fig.\,1, $A=1.09\times 10^{7}$\,C$^{-2}$m$^{2}$NK$^{-1}$,
$b\approx 0$; $c=3.3\times 10^{14}$\,C$^{-6}$m$^{10}$N, 
$G=8.26\times 10^{-10}$\,C$^{-2}$m$^{4}$N, the vacuum dielectric permittivity
$\varepsilon _{0}=8.85\times 10^{-12}$\,C$^{2}${m}$^{-2}${N}$^{-1}$, and the background non-critical part of the dielectric constant 
$\varepsilon_i=20$.
The encountering of oncoming antiparallel domains is achieved by the special choice of boundary conditions for polarization at the top and at the bottom of the sample, see Methods.

Figure\,1b  displays the simulated structure of the interpenetrating domains with the respective up- and down- polarization directions shown by white arrows. 
The fair correspondence between the experimentally observed, Fig.\,1a, and simulated, Fig.\,1b, domain textures is clearly seen. 

To display the spatial polarization distribution, we make an inner-corner cut of the simulated PGO sample, see Fig.\,1b.  
The structure reveals the self-similar configuration with branching domains. 
 The vertical cross-cuts demonstrate the inclined domain walls, the cuts crossing the meanders producing the famed ``cigar-shaped" forms.  
 The horizontal cross-cut exposes the irregular pattern of cross-sections of the domains with the DWs highlighted by green lines. Another important feature is that the upper and bottom near-surface regions host the Landau fractal multidomain structure that forms to compensate surface bound charges\cite{Strukov2012,Guerville2005}. 

To summarise here, both experimental data and simulations establish the existence of the bent DWs, where, supposedly, substantial bound charges, destabilizing ferroelectricity should arise. 
At the same time, the electric charge, $\rho=-\mathrm{div}\,\mathbf{P}$, appears to be vanishingly small in the vicinity of the DWs ridges, where it is supposed to be the maximal 
(see the charge map of the horizontal cross-section area, magnified at the lower plane of the Fig.\,1c). 
The bound charge at the bent DWs is indeed much smaller than the charge  expected from the $z$-gradient of the spontaneous polarization, $\rho_z=-\partial_z
P_z$, presented at the upper plane of Fig.\,1c; the up– and down- arrows in the upper plane show the directions of the $P_z$ polarization. 
The solution of the puzzle is that the $\rho_z$ charge is nearly compensated by the shown in the middle plane charge $\rho_{xy}=-\partial_x P_x - \partial_y P_y$ induced by the transverse, $P_x$ and $P_y$, polarization components, that effectively screen the charges $\rho_z$; the in-plane arrows in the middle plane indicate the directions of the transverse polarization.  
Hence, the total bound charge arising near the bent DWs, $\rho=\rho_{z}+\rho_{xy}$, comes out much smaller than $\rho_z$. 
The more detailed distribution of the polarization fields and charges is presented in Supplementary Fig.\,2.  
 This figure also reveals the softening of the H-H and T-T DWs resulting in their broadening  with respect to the vertical uncharged DWs which occurs due to the interplay between the polarization and electric field induced by the bound charges as predicted in\,\cite{Guerville2005,Lukyanchuk2009}.

The obtained smallness of the bound charge at DWs, supports the basic concept that uniaxial ferroelectrics maintain the divergenceless character of the polarization. Indeed, Fig.\,1d displays the polarization distribution as flux bundles visually identical to the flowing towards each other and clash-avoiding streams of the incompressible liquid. Accordingly, the polarization forms a network of the counterflowing slightly bent domains alternatively carrying ``up" and ``down" polarization fluxes. 
It is the transverse component of the polarization that ensures that the domains with opposite flux avoid the mutual clash.

The counterflowing domains are separated by domain walls as shown in detail in Fig\,2.
Figure\,2a presents an essential building block of the domain network, 
the two branching streams assuming the V-shape clash-avoiding scissoring  configuration.  
The global image of the domain wall is shown in Fig.\,2b and constitutes a path-connected topological manifold. Figure\,2c shows the side view of the DWs manifold. Figure\,2d is a tilted view clearly exposing the scissoring point $\mathbb{V}$ marked red. 
This point realizes a saddle node singularity of the polarization vector field $\mathbf{P}$ (Fig.\,2e). 
The node is characterized by the singularity index calculated as the 
sign of the Jacobian of the field $\mathbf{P}$ at point $\mathbb{V}$,  $N=\mathrm{sgn}\,\mathrm{det} \left\Vert \partial _{i}P_{j}\right\Vert$\,\cite{Dubrovin2012}. 
Accordingly, $N$ is equal to $+1$ when the outcoming in the $\pm$z-direction polarization fluxes of T-T encountering domains captivate the polarization flow converging to the $\mathbb{V}$-point from the x-y plane; this situation is shown in Fig.\,2e. In the opposite case of the H-H domains encountering, $N=-1$.

\begin{figure*}[h!]
\center
\includegraphics [width=18cm] {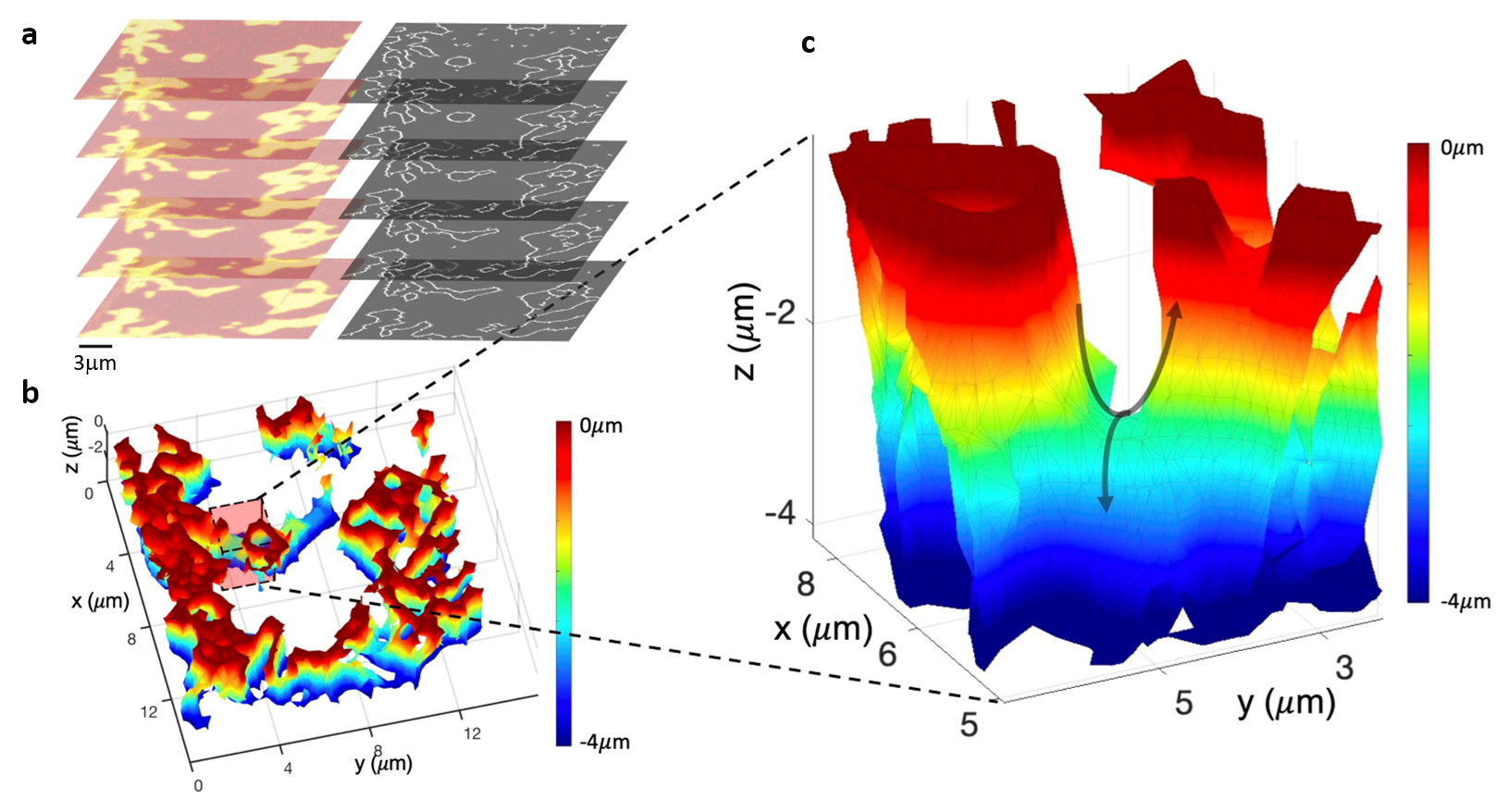}
\vspace{1cm}
\caption{ \textbf{Domain wall saddle points in lead germanate} 
\textbf{a,}\,Vertical PFM phase (left) of lead germanate at different depths into the crystal, obtained by PFM tomography, along with the domain wall trace inferred from the PFM (right). The scale bar is 2µm. 
\textbf{b,}\,3D domain wall reconstruction, obtained by triangulating the points identified as the the DW surface in \textbf{b}. 
\textbf{c,}\,A zoomed in view of the domain wall saddle point highlighted in \textbf{c}. Arrows show the intersecting maxima and minima, and the color scale represents z height.
}
\label{FigPGOexperiment}
\end{figure*}

The singularity of the field at point $\mathbb{V}$ is a topological property. To see that, let $S^2$ be a closed sphere centered at $\mathbb{V}$ that is pierced by the polarization field $\mathbf{P}$. Consider now the map of $S^2$ onto the space of the tips of the normalized vector $\mathbf{n}=\mathbf{P}/P$ which is also a 2-dimensional sphere ${S^\prime}^2$. Thus, each point $\in S^2$ has an image on  ${S^\prime}^2$ corresponding to the vector $\mathbf{n}$ at this point. This correspondence is characterized by the degree of the map $S^2\overset{f}{\rightarrow} {S^\prime}^2$\,\cite{Dubrovin2012},  
\begin{equation}
\mathrm{deg}\,f=\frac{1}{4\pi }\int_{S^2}\left( \mathbf{n}\cdot\left[ \partial _{\theta}\mathbf{n}\times \partial _{\varphi} \mathbf{n} \right] \right) {d\theta d\varphi}
\label{ind}
\end{equation}
where the integration is performed over the surface $S^2$, with $\varphi$ and $\theta$ being the spherical coordinates. 
This quantity is an integer number giving the number of times that the image of sphere $S^2$ is wrapped over the sphere ${S^\prime}^2$, hence it is the topological characteristic of the point $\mathbb{V}$ that does not depend on the choice of the wrapping surface. Importantly, the degree of the map $S^2\overset{f}{\rightarrow} {S^\prime}^2$ for the sphere surrounding the point $\mathbb{V}$ is identical to the singularity index $N$ for this point,  $\mathrm{deg}\,f=N$. This fundamental statement establishes the deep relation between the topological properties of the point $\mathbb{V}$ and the differential properties of the polarization vector field, $\mathbf P$, in its vicinity. Therefore, the stability of the domain scissoring structure hosting the singular saddle point $\mathbb{V}$ is topologically protected.

It is clearly seen that in the close vicinity of the saddle point $\mathbb{V}$, the shape of the DW is the saddle surface i.e., is a surface that curves up and curves down in perpendicular directions. The color of the manifold, see Fig.\,2d, represents the effective surface charge density at the manifold which remains indeed low. 
The DW assumes the configuration that minimizes its energy under constraints imposed by the boundary conditions stemming from the domains' collisions. The shape of the DW 
is similar to that of the soap film, which is attached to the outer fixed frame and shapes itself to minimize its surface energy\,\cite{Maggi2018}. The DW is thus an exemplary realization of a concept widely known in mathematics as the Lagrange minimal surface problem\,\cite{Dierkes2010}.
Note that our multiconnected domain wall presents a more general case than the usual minimal surface
of the soap film: it accounts also for the minimization of the non-local energy of bending of the DW-generating counter-flowing polarization field lines.

One of the key characteristics of this DW surface is the presence of an array of the saddle
points seen clearly in Fig.\,2b. Such saddle points, however cannot be detected using conventional 2D PFM domain
mapping. Fortunately, recent advances in tomographic
imaging  genuinely render high-resolution 3D images of ferroelectric microstructures\,\cite{Steffes2019,Kirbus2019} to access the 3D domain wall distribution in the PGO. Here we use the tomographic atomic force
microscopy (TAFM)\,\cite{Steffes2019}. The TAFM involves
rastering the tip of a scanning probe microscope, using significant compressive force, as high as
microNewtons, to sequentially mechanically remove thin layers from the specimen surface.
Imaging can be done either at the same time as the machining (and under the same conditions), or
in separate scans taken between machining runs, in which more conventional tip pressures can
be applied for imaging. We take the former approach. Machining and imaging in the same scanned
area, allows for acquiring a series of planar domain images at different depths into the sample to be collected,
as illustrated in Fig.\,3a. Here the vertical PFM phase maps, taken at differing depths below a polished
PGO polar surface, are stacked alongside equivalent images in which the DWs are highlighted using image processing techniques in which maxima in colour gradients are
plotted. 

These images enable rendering of the DW surface in 3D (Fig.\,3b).
Several important observations are made: firstly, the DW appears as a single
continuous meandering surface throughout most of the investigated area; secondly, the DW
surface contains multiple saddle points. The pink box in Fig.\,3b highlights the most
obvious one found in this dataset. This region is isolated and magnified for clarity in Fig.\,3c; additional saddle points are explicitly highlighted in the Supplementary Fig.\,3. 


Our results conclusively establish the existence of a new topological state of matter, corresponding to a counterflowing domain state where the DW separating clash-avoiding polarization counterflows is a unique multiconnected surface. This clash-avoiding structure characterized by a multitude of the saddle points at the DW surface allows to avoid formation of the uncompensated charges at the locations of the H-H and T-T polarization flow impacts
and maintains the electric neutrality of the interior of the ferroelectric material.  Our discovery of almost uncharged H-H and T-T regions in PGO is entirely consistent with the observation that the domain walls in PGO do not manifest enhanced conductivity, unlike systems in which bound charges emerging at DWs are screened by the intrinsic current-conducting carriers. 
The reported results validate the introduced topological concept describing ferroelectrics in terms of incompressible fluids.

\newpage

\section*{Methods}~~\newline
\subsection{Tomographic atomic force microscopy (TAFM).}
The TAFM experiments (Fig.\,3) were performed on the polar face of the same lead germanate crystal, using an Asylum Research MFP-3D Infinity AFM system with an internal lock-in amplifier. The TAFM technique employed involves simultaneous PFM imaging of the domain structure and AFM machining. A conductive diamond-coated silicon probe was scanned over the surface at a speed of 75\,$\mu$ms$^{-1}$. The probe has a high stiffness constant of 80\,Nm$^{-1}$, which, when combined with a high AFM deflection setpoint, produces a significant force on the sample surface. This resulted in the sequential removal of layers of material at a rate of approximately 0.2\,$\mu$m per scan, up to a total depth of ~4\,$\mu$m, over a 20\,$\mu$m $\times$ 20\,$\mu$m region. Domain information was collected simultaneously via PFM (AC bias amplitude 2V and frequency ~3.0 MHz) allowing for the domain evolution through the depth of the crystal to be mapped. The locations of the domain walls at various depths were extracted by taking the gradient of the PFM phase maps collected during TAFM. Using these domain wall locations, a 3D reconstruction of the domain wall surface (Fig 3a) was created using the “alphaShape” triangulation function, implemented in MATLAB. All PFM and TAFM data were collected at room temperature under ambient conditions. 

\medskip

\noindent \textbf{Phase-field method}.
    To perform numerical calculations we used phase-field method implemented within FEniCS software package\,\cite{LoggMardalEtAl2012a}. Nonlinear system of partial differential equations to solve are coming from the variation of free-energy functional (\ref{GL}) with respect to polarization $P$:
    \begin{equation}
        -\gamma \frac{\partial P}{\partial t} = \frac{\delta F}{ \delta P},
    \label{FunctionalVariation}
    \end{equation}
where $\gamma$ is the time-scale parameter, which value is irrelevant for current case and is taken to be equal unity. 
    The nonlinear system of equations (\ref{FunctionalVariation}) is closed by the linear system corresponding to electrostatic Poisson equation:
    
    \begin{equation}
        \varepsilon_{0} \varepsilon_{i} \nabla^{2} \varphi = \partial_z P,
    \label{Poisson}
    \end{equation}
    where $\varphi$ is the electrostatic potential.
    
    The computational domain is a rectangular box with dimensions $200 \times 200 \times 500$ nm, that was partitioned into tetrahedrons with the help of 3D mesh generator gmsh \cite{Geuzaine2009}.
    At the top and bottom surfaces of the computational domain the Dirichlet boundary conditions are imposed, $P_{top} = P^{\prime}\mathrm{sin}(2\pi x / 100 + \pi) \mathrm{sin}(2\pi y / 100)$, $P_{bottom} = P^{\prime}\mathrm{sin}(2\pi x / 100) \mathrm{sin}(2\pi y / 100)$, where $P^{\prime}=5$\,$\mu$C\,cm$^{-2}$, the value close to the equilibrium value of polarization throughout the volume of sample. The phase shift between polarization distributions at the top and bottom rectangular surfaces was properly selected to ensure encountering of the oppositely polarized domains in the bulk of the sample. In the $x$ and $y$ directions the boundary conditions on $P$ and $\varphi$ are set to be periodic.
    
    Approximation of time derivative on the left hand side of variation (\ref{FunctionalVariation}) is accomplished by BDF2 stepper with variable time step \cite{Janelli2006}. 
    At the first time step of the simulation, the values of the polarization in the computational domain are taken randomly from the $[-10^{-4},10^{-4}]$\,$\mu$C\,cm$^{-2}$ interval.
    Solution of nonlinear system of equations is performed with the Newton based nonlinear solver with line search and generalized minimal residual method with restart\,\cite{petsc-web-page,petsc-user-ref}. Linear system resulting from discretization of Poisson  equation (\ref{Poisson}) is solved using the generalized minimal residual method with restart.

\section*{Acknowledgements}~~\newline
This work was supported by H2020 ITN-MANIC action (Y.T, J.M.G , I.L.). J.M.G., C.J.McC. and J.P.V.McC., acknowledge funding from the Engineering and Physical Sciences Research Council (grant number EP/P02453X/1 and PhD studentship support). J.M.G and J.R.M. acknowledge studentship support from the Northern Ireland Department for the Economy (DfE).  Research at
the University of Nebraska (A.G.) was supported by the National Science
Foundation (NSF) grant DMR-1709237. The work of V.M.V. was supported by Fulbright Foundation and by Terra Quantum AG.


\renewcommand{\figurename}{Supplementary Figure}
\setcounter{figure}{0}    

\begin{figure*}
\center
\includegraphics [width=11 cm] {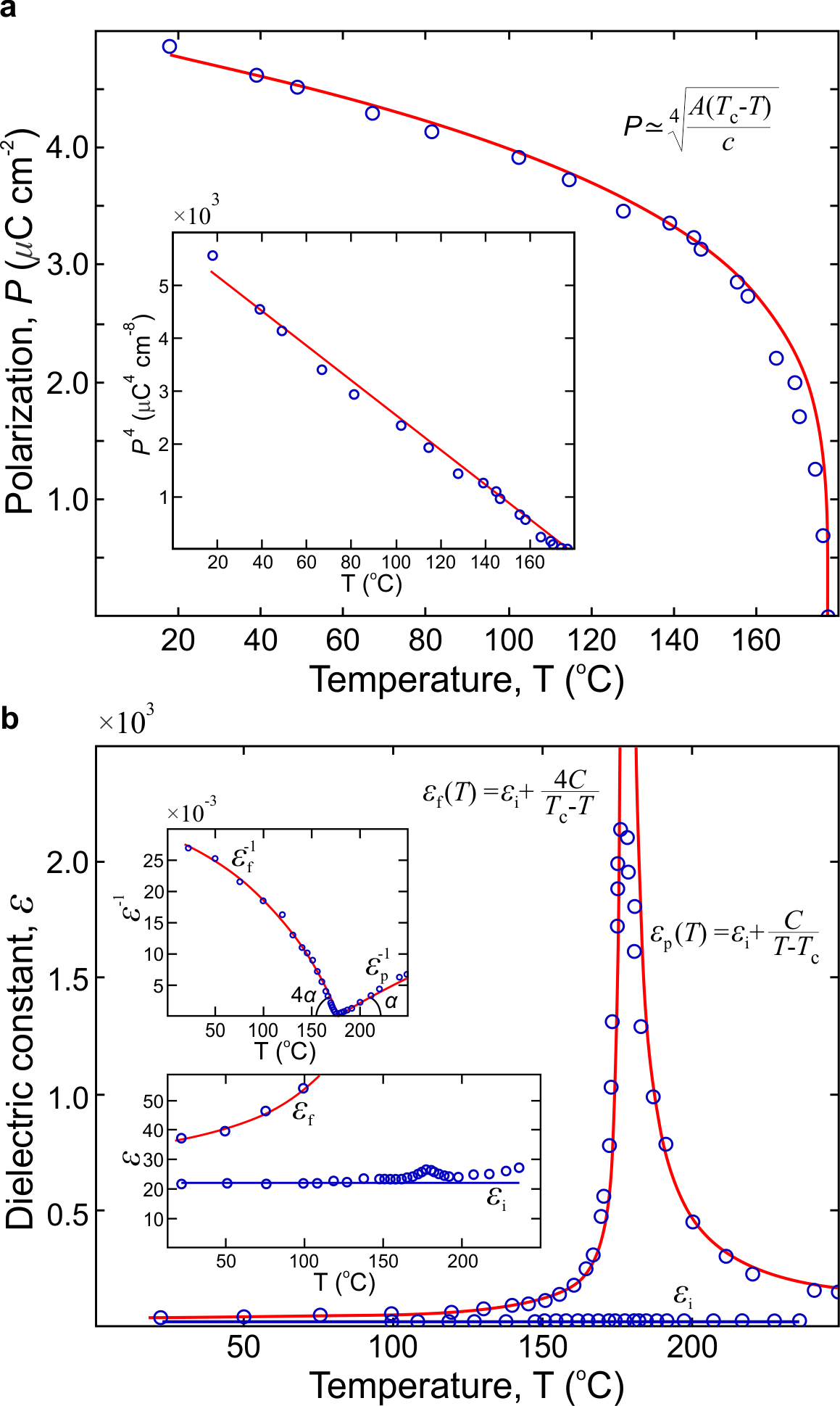}
\caption{ \textbf{Determining Ginzburg-Landau functional parameters from the experimental data}\,\cite{Iwasaki1972}. 
\textbf{a,}\,Fitting the experimental data (blue circles) of the polarization temperature dependence  reveals the critical exponent $1/4$ below $T_c$ which means that the coefficient $b$ in the functional (1) of the main text is anomalously small and the system is close to the tricritical point where order of the transition is changed from the 2nd to the 1st one\,\cite{Strukov2012}. The inset showing linear dependence of $P^4$ on $(T_c-T)$ offers a strong support to this fit. 
\textbf{b,}\,Fitting the experimental temperature dependence of the dielectric constants, $\varepsilon_f$ and $\varepsilon_p$ of the ferroelectric and paraelectric phases, measured in the $z$-direction, reveals the Curie-like behaviour. The Curie constant at the paraelectric side  $C=10400$\,K of the transition has the magnitude four times smaller then its magnitude at the ferroelectric side. This feature again reflects that the system is close to the tricritical point with $b\approx 0$. The upper inset, demonstrating the temperature dependence of the inverse dielectric constant, strongly confirms this relation as the angles between the tangents to $\varepsilon_f^{-1}(T)$ and $\varepsilon_i^{-1}(T)$ and temperature axis at $T=T_c$ differ by factor of 4. The downward bending of the $\varepsilon_f^{-1}(T)$ dependence at low temperatures is due to the background dielectric constant that is found to be equal $\varepsilon_i=21$. The bottom inset uses the magnified $\varepsilon$ scale to highlight the nearly temperature-independent behavior of the transversal dielectric constant appearing to be equal to $\varepsilon_i$, which is barely seen in the main plot. Identification of the Curie constant $C$ allows for determining parameter $A=(\varepsilon_0 C)^{-1}=1.09\times 10^{7}$\,C$^{-2}$m$^{2}$NK$^{-1}$ in functional (1) of the main text, which, in turn, allows for finding parameter $c=3.3\times 10^{14}$\,C$^{-6}$m$^{10}$\,N from the fit of the polarization dependence of panel\,\textbf{a}. The coefficient $G=8.26\times 10^{-10}$\,C$^{-2}$m$^{4}$N in the gradient term of functional (1), which is assumed to be isotropic, is estimated by taking the coherence length  $\xi_0=[G/(T_cA)]^{1/2}$ of the order the elementary cell size 0.4\,nm.
}
\label{FigCurves}
\end{figure*}


\begin{figure*}
\vspace{2cm}
\center
\includegraphics [width=20 cm] {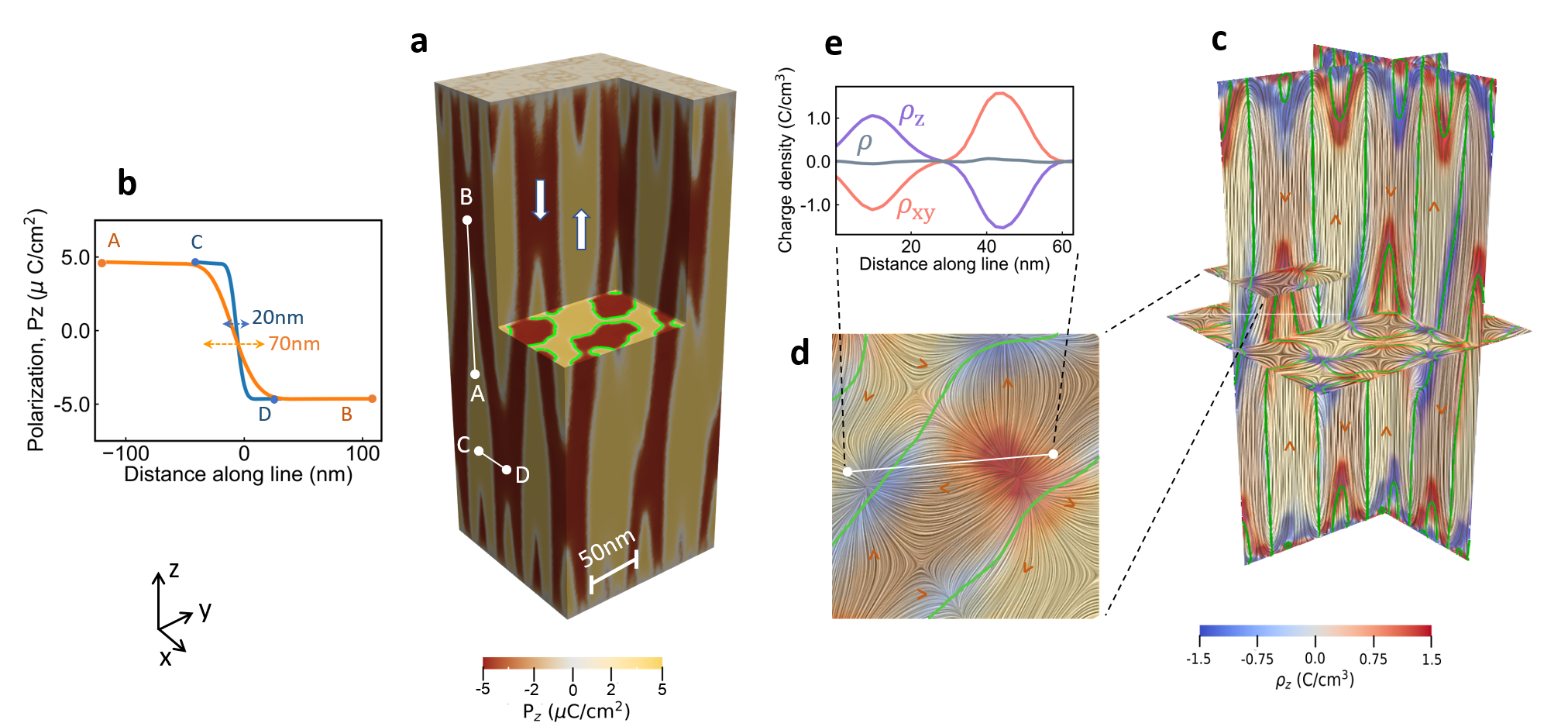}
\caption{ \textbf{Polarization and charge distribution }. 
\textbf{a,}\,\,A sketch of the inner-corner cut of the rectangular uniaxial PGO sample reproducing Fig.\,1a of the main text visualizing the simulated domain configuration.
The panel displays the structure of the interpenetrating polarization domains with the respective up- and down- polarization directions shown by white arrows. The magnitude of the z-component of the polarization, $P_z$, is quantified by the colour legend. The upper and bottom near-surface regions display a Landau fractal multidomain branching structure which arises to compensate the surface bound charges. The green contours in the horizontal plane of the corner cut, depict the cross sections of the domain walls. The AB and CD lines represent the along- and across-domain cuts along which the distribution of the polarization component $P_z$ is taken.
\textbf{b,}\,The sketch of the distribution of the spontaneous polarization along the AB and CD cuts. Importantly, the profile of the of the charged DW arising in the head-to-head (H-H) and tail-to-tail (T-T) polarization collisions along the AB is more broader than the corresponding profile uncharged domain wall along CD. The softening of the H-H and T-T domain wall occurs due to the interplay between the polarization and electric field induced by the bound charges as was predicted in\,\cite{Guerville2005,Lukyanchuk2009}.  
\textbf{c,}\,The 3D map of the polarization and bound charge distribution. The lines represent the flow of the polarization field and the spacial color map shows the corresponding distribution of the charges produced by the spontaneous polarization, $\rho_z=-\partial_z P_z$. The red triangles mark the direction of the polarization field. 
\textbf{d,}\,The distribution of the polarization lines and the bound charge in the magnified  xy-plane that screens the spontaneous bound charges $\rho_z$. 
\textbf{e,}\,The illustration of the screening effect. The spontaneous bound charges $\rho_z=-\partial_z P_z$ are almost completely balanced by bound charges $\rho_{xy}=-\partial_x P_x-\partial_y P_y$ produced by the transversal polarization field with components $P_x$ and $P_y$, so that the resulting charge $\rho$ nearly vanishes.
}
\label{FigCurves}
\end{figure*}


\begin{figure*}
\vspace{1cm}
\center
\includegraphics [width=14 cm] {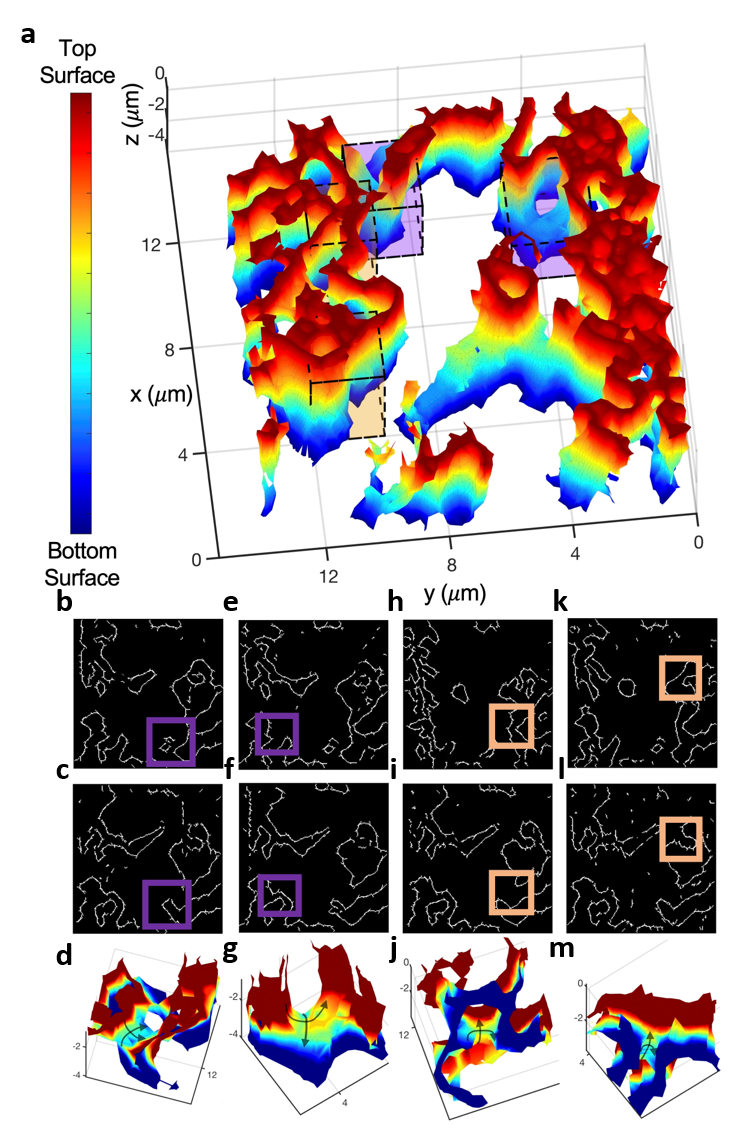}
\caption{ \textbf{Domain wall saddle points}. 
\textbf{a,}\,PGO domain wall reconstruction. The color bar represents $z$-height below the surface, and the boxed regions indicate further saddle points.
\textbf{b-c,}\,Domain wall trace extracted from the PFM tomography data. \textbf{b} occurs towards the top surface of the sample, and \textbf{c} is obtained after some milling, at a greater depth. The boxed region shows where domain walls merge to produce a saddle point. 
\textbf{d,}\,The saddle point corresponding to the box in \textbf{b-c}, as represented by the 3D domain wall recreation. 
\textbf{e-g, h-j} and \textbf{k-m}  follow the same pattern, for different saddle points.
Throughout the figure, purple boxes correspond to saddle points visible from the top down (where domains merge with increasing depth), and orange boxes correspond to saddle points visible from the bottom up (domains are initially merged and then separate with increasing depth).
}
\label{FigCurves}
\end{figure*}


\clearpage
\printbibliography 


\end{document}